\newcommand{\diracslash}[1]{#1\llap{/\kern2pt}}

\newcommand{\be}{\begin{equation}}
\newcommand{\ee}{\end{equation}}
\newcommand{\bea}{\begin{eqnarray}}
\newcommand{\eea}{\end{eqnarray}}
\newcommand{\ba}[1]{\begin{array}{#1}}
\newcommand{\ea}{\end{array}}

\documentclass[prd,aps,floats,nofootinbib,tightenlines,showpacs]{revtex4}
\usepackage{epsfig,graphicx,pstricks}
\usepackage{psfrag}
\usepackage{color}
\usepackage{amsmath}
\usepackage{amsfonts}
\usepackage{amssymb}
\usepackage{color, soul}
\usepackage{textcomp}
\usepackage{multirow}
\usepackage{natbib}
\usepackage{subfigure}
\begin{document}
\preprint{APS/123-QED}
\title{Study of finite volume number density fluctuations in the SU(3) Polyakov loop extended Nambu–Jona-Lasinio model for the search of the QCD Critical Point}% Force line breaks with \\
\author{A.~Sarkar}
\email{amal.sarkar@cern.ch}
\author{P.~Deb}
\email{paramita.dab83@gmail.com}
\author{R.~Bose}
\email{rohannewone@gmail.com}
\author{R.~Varma}
\email{varma@phy.iitb.ac.in}

\affiliation{*School of Physical Science, Indian Institute of Technology Mandi, Kamand, Mandi - 175005, India}
\affiliation{†\S Department of Physics, Indian Institute of Technology Bombay, Powai, Mumbai- 400076, India}
\affiliation{{‡}Department of Physics, Ramakrishna Mission Residential College (Autonomous), Narendrapur, Kolkata - 700103, India}

%\affiliation{
%\affiliation{
 %Third institution, the second for Charlie Author
\begin{abstract}
The critical endpoint (CEP) is a fundamental feature of the Quantum Chromodynamics (QCD) phase diagram, marking the boundary between quark-gluon plasma and hadronic matter. Heavy-ion collision experiments, such as the RHIC Beam Energy Scan, aim to probe the QCD phase diagram by varying collision energy. However, the short-lived nature of produced particles makes direct measurements challenging, necessitating theoretical models. This study explores the impact of density fluctuations on the CEP using the Polyakov-loop enhanced Nambu–Jona-Lasinio (PNJL) model, focusing on quark number densities in both finite and infinite volume systems. Quark number densities, derived from thermodynamic susceptibilities, serve as reliable predictors for the CEP’s location. We calculate density fluctuations and normalize them by $T^3$ as functions of temperature and $T/\mu_x$ (where x represents light quarks, strange quarks, and baryons), analyzing inflection points and maxima to estimate the critical region. To compare the experimental data, the study has been performed at energies identical to those of the RHIC Beam Energy Scan. The results highlight the influence of finite volume effects on quark density fluctuations, key indicators of QCD phase transitions, and provide quantitative comparisons with experimental data. This work enhances our understanding of QCD phase structure and supports the ongoing search for the CEP in high-energy heavy-ion collisions, bridging theoretical predictions and experimental observations.

\end{abstract}
\maketitle
%\tableofcontents
\section{\label{sec:level1}Introduction%\protect\\ The line 
%break was forced \lowercase{via} \textbackslash\textbackslash
}
In our endeavor to study the early universe, we must understand the matter that existed at that time. If we can reproduce this matter under laboratory conditions, matching the density and temperature of that era, we could study the characteristics of the early universe. The only feasible method so far is to collide two heavy atomic nuclei, accelerated to energies on the order of a hundred GeV, in a collider. The Relativistic Heavy Ion Collider (RHIC) at Brookhaven National Laboratory (BNL) and the Large Hadron Collider (LHC) at CERN are the largest colliders for this purpose. Head-on collisions in these colliders provide conditions similar to the density and temperature that existed when the universe was first formed. Over the past few decades, physicists have been attempting to recreate quark-gluon plasma (QGP) by colliding atomic nuclei at sufficiently high energies, producing temperatures of trillions of degrees. At these high temperatures, matter exists in the QGP state for a brief moment. When two particle beams collide head-on, some of the matter transitions from a hadronic state to a QGP state. During this chiral phase transition, a first-order phase transition may occur. The possible existence of a first-order phase transition and a smooth crossover implies that the transition line should terminate at a point, known as the QCD critical point or critical end point (CEP)  {\cite{STAR, STAR1}}. The CEP marks the end of the first-order line, where the phase transition becomes second-order. The critical point for the QCD phase diagram is where the boundary between phases disappears. To locate the critical point in the phase diagram, physicists study fluctuations related to observables of final state particles, such as pion multiplicity or the kaon-to-pion ratio, as well as higher-order moments of conserved quantities. When a heavy-ion collision occurs at sufficient energy, quark-gluon plasma is formed in a finite volume. This volume depends on the size of the colliding nuclei, center-of-mass energy, and collision centrality. Studies on different centrality measurements of Hanbury-Brown-Twiss (HBT) radii suggest that the system volume increases with centrality during freeze-out \cite{adamova}. Several theoretical models, such as the non-interacting bag model \cite{elze}, chiral perturbation theory \cite{luscher, gasser}, NJL \cite{nambu, kiriyama}, PNJL model \cite{fukushima,ratti,pisarski,fukushima1,hansen,ciminale,ghosh,deb1,abhijit,bhattacharyya2010investigation,deb2009mesonic}, and by the first principle study of pure gluon theory on space-time lattices \cite{bazavov,boyd,engels,fodor,allton,forcrand,aoki,megias} have examined the effects of finite volume on the QGP system. In a $1+1$ dimensional NJL model, the finite size effects of dense baryonic matter have been explored through the induction of charged pion condensation. This study has recently been extended to the Polyakov loop Nambu–Jona-Lasinio (PNJL) model, where it was observed that as the system’s volume decreases, the temperature at the critical point also decreases, causing the CEP to shift to a region with higher chemical potential $(\mu)$ and lower temperature $(T)$ {\cite{deb1, amal1, abhijit}}. More generally, both lattice QCD calculations and QCD-based models suggest that fluctuations in strongly interacting matter at zero density exhibit significant volume dependence, which may have implications for understanding fireball formation in heavy-ion collisions. In the PNJL model, it has been observed that as volume decreases, the critical temperature for crossover transitions also decreases. For lower volumes, the CEP shifts to regions of higher chemical potential and lower temperature. These findings indicate that the CEP’s location strongly depends on the size of the fireball produced in heavy-ion collisions.

Identifying the CEP of the QCD phase diagram provides insight into the critical point of the early universe, specifically the temperature and state of matter just after the big bang. Over time, various approaches have been developed to locate the CEP. With technological advancements, we are no longer limited to theoretical models but also have experimental possibilities. Heavy-ion colliders are crucial in this experimental search, where ions, protons, and other particles have collided to study their fundamental constituents. However, since these collisions are instantaneous and the produced particles are short-lived, it is difficult to observe them directly, necessitating models to estimate the observables. An important development in this exploration was the proposal for expanding the chiral Lagrangian, such as the NJL model, by connecting quarks to a uniform temporal background of the gauge field (the Polyakov loop). The Polyakov-Nambu-Jona-Lasinio (PNJL) model is one such effective model that demonstrates two essential characteristics of QCD, which are spontaneous chiral symmetry breaking and confinement-like properties.

In this study, we investigate the effect of density fluctuations on the critical endpoint of the QCD phase diagram using the PNJL model. We focus on calculating quark number densities and attempt to locate the critical region, comparing our findings with the results from the experimental RHIC Beam Energy Scan program \cite{tlusty2018rhic,odyniec2019beam}. It is known that quark number densities are important thermodynamic quantities, as quark number susceptibilities at finite temperatures and densities are derived from the quark number density’s derivatives with respect to the chemical potential \cite{zong}. Since quark number susceptibilities are good indicators for predicting the critical endpoint, quark number densities should also provide a reliable prediction for the CEP’s location. Thus, investigating the number densities within the PNJL model is vital for exploring the QCD critical point.  This work focuses on exploring the influence of finite volume effects on quark number density fluctuations, which are believed to be key signatures of the phase transition. The study aims to provide deeper insights into the phase structure of QCD and to contribute to the ongoing search for the elusive critical end point in high-energy heavy-ion collisions. We briefly discuss the formalism of the PNJL model in section 2. In the subsequent section, we estimate the location of the critical endpoint. We present the number density and number density divided by $T^3$ as a function of temperature and $T/\mu_x$ (where x represents light quarks, strange quarks, and baryons). By analyzing the curves, we identify points of inflection and maxima, which provide an estimate of the critical region and the CEP of the QCD phase diagram.

\section{\label{sec:level1}PNJL Model}
Effective models are fundamental in studying high energy physics phenomena. They are constructed with Lagrangians of a simple mathematical structure having properties similar to that of QCD. Spontaneous chiral symmetry breaking, global Z(3), and confinement-like properties, which are crucial characteristics of QCD have been demonstrated by the PNJL model which is made possible due to the symmetry of its Lagrangian. 
PNJL is an extension of the Nambu-Jona-Lasinio model, that can serve this purpose. The Polyakov loop in the PNJL model represents the interactions between quarks with the temporal gluon field. One of the important development in the formalism of the model was the expansion of the quark chiral Lagrangian to a gauge field with uniform temporal background. In this model, the Polyakov loop dynamics serve as a framework for gluon dynamics representing chiral point couplings between quarks and a background. In PNJL, a non-local interaction is used instead of a point-like four-fermion interaction. This model is such an effective model that gives good results at finite temperature and chemical potential. In this study, we have considered the 2+1 flavor PNJL model with six quark interactions. In the PNJL model, chiral point couplings between quarks (present in the NJL portion) and a background gauge field that represents Polyakov Loop dynamics are used to describe gluon dynamics. The Polyakov loop is written as,
\begin {equation}
  L(\bar x)={\cal P} {\rm exp}[i {\int_0}^\beta
d\tau A_4{({\bar x},\tau)}]
\end {equation}
where the temporal component of Eucledian gauge field $(\bar A,A_4)$ is represented by $A_4=iA_0$. $\beta=\frac {1}{T} $ with Boltzmann constant $K_B$ put to one, and $\cal P$ indicates path ordering. Under global $Z(3)$ symmetry, the Polyakov line $L(\bar x)$ turns into a field with a charge of one. The Polyakov loop field is then introduced by,
\begin{equation}
\begin{split}
\Phi=(Tr_{_c}L)/N_{_c} \\
\text{and\quad it\'s\quad conjugate\quad} \bar{\Phi}=(Tr_{_c}L^{\dag})/N_{_c}
  \end{split}
\end{equation}
The three flavor of the Lagrangian of the PNJL model and $U_A(1)$ anomaly may be written as :
%\begin{multiline}
\begin{equation}\label{eqn.1}
\begin{split}
%\begin{eqnarray}
%\begin{aligned}
 \cal{L}&=\sum_{f=u,d,s}\overline{\psi}_f(i\gamma^{\mu}D_\mu- m_f +\gamma_0\mu)\psi_f + \frac{g_S}{2} \sum_{a=0}^8[(\overline{\psi}\lambda^a\psi)^2\\
  & +(\overline{\psi}i\gamma_5\lambda^a\psi)^2]-g_D\{det[\overline{\psi}_f\frac{(1+\gamma_5)}{2}\psi_{f^\prime}]+\\
  & det[\overline{\psi}_f\frac{(1-\gamma_5)}{2}\psi_{f^\prime}]\} -\mathcal{U^\prime}(\Phi[A],\overline{\Phi}[A];T)
%\end{aligned}
%\end{eqnarray}    
\end{split}
\end{equation}
%\end{multiline}
where $D_\mu=\partial_\mu-iA_\mu$ and $A_\mu$ is the gauge field that absorbs the strong interaction coupling and is given by $A_\mu(x)=g_S A_\alpha^\mu\lambda^\alpha/2$. $\lambda^a$ are the eight Gell-Mann matrices. $f$ symbolises the flavors $u$, $d$ or $s$ respectively. The three-flavor current quark mass matrix is given by $m_f$, in which $m_f$ = diag $(m_u, m_d, m_s)$. For two flavors $g_D$ is equal to zero. The left-handed and right-handed chiral projectors are respectively given by the matrices $P_{L,R}=(1\pm \gamma_5)/2$, and the other terms have their usual meaning, described in detail in Refs.~\cite{bhattacharyya2010investigation,deb2009mesonic}. The confinement/ deconfinement properties of the quarks are described by the effective potential $\mathcal{U}^\prime (\Phi,\bar{\Phi};T)$ (with the Vandermonde term $J[\Phi,\overline{\Phi}]=(27/24\pi^2)[1-6\Phi\overline{\Phi}+4(\Phi^3+\overline{\Phi}^3)-3(\Phi\overline{\Phi})^2]$) demonstrated as:
\begin{align*}
    \frac{\mathcal{U}^\prime(\Phi[A],\overline{\Phi}[A];T)}{T^4}=\frac{\mathcal{U}(\Phi[A],\overline{\Phi}[A];T)}{T^4}-\kappa ln[J(\Phi,\overline{\Phi})]
\end{align*}

where $\mathcal{U}(\phi)$ is a Landau-Ginzburg type potential expressed as,
\begin{equation}
   \frac{ \mathcal{U}(\Phi,\overline{\Phi};T)}{T^4}=-\frac{b_2(T)}{2}\overline{\Phi}\Phi-\frac{b_3}{6}(\Phi^3+\overline{\Phi}^3)+\frac{b_4}{4}(\overline{\Phi}\Phi)^2
\end{equation}
with $b_2(T)=a_0+a_1\left(\frac{T_0}{T}\right)+a_2\left(\frac{T_0}{T}\right)^2+a_3\left(\frac{T_0}{T}\right)^3$. For the effective potential, we can choose the following parameters: $T_0 =0.19$ GeV, $a_0 =6.75, a_1 =-
1.95, a_2 =2.625, a_3 =-7.44, b_3 =0.75, b_4 =7.5, \kappa=0.13$.
\par
Now, we know that the energy of a quark can be written as, $E_i = \sqrt{ p_i^2 +m_i^2}$. Using this and Eq.\,(\ref{eqn.1}), we can obtain the Grand potential density for the PNJL model in MFA (mean field approximation) as,
\begin{equation}
\begin{split}
  \Omega^{\prime}(\Phi,\bar{\Phi},\sigma_{_f},T,\mu_{_f})=u^{\prime}[\Phi,\bar{\Phi},T]+
  2g_{_s}\sum_{{f=u,d,s}}(\sigma^{2}_{_f}-\frac{g_{_D}}{2}\sigma_{_u}\sigma_{_d}\sigma_{_s})
  \\-T\sum_{_n}\int^{\infty}_{_\lambda}\frac{d^{3}p}{(2\pi)^{3}}Tr ln\frac{S^{-1}(i\omega_{_n},\vec{p})}{T}
\end{split}
\end{equation}
where $\omega_{_n}=\pi T(2n+1)$ are Matsubara frequencies for fermions. The inverse quark propagator in momentum space is given by
\begin{equation}
  S^{-1}=\gamma_{_0}(p^{0}+\hat{\mu}-iA_{_4})-\vec{\gamma}.\vec{p}-\hat{M}.
\end{equation}
Using the identify Tr ln(X) = ln det(X), we get

\begin{widetext}
\begin{equation}
    \begin{split}
        \Omega^\prime &=\mathcal{U}^\prime[\Phi, \overline{\Phi}, T] + 2g_S \sum_{f=u,d,s} \sigma_f^2-\frac{g_D}{2}\sigma_u\sigma_d\sigma_s-6\sum_f\int_0^\Lambda \frac{d^3p}{(2\pi)^3}E_{pf}\Theta(\Lambda-|\Vec{p}|)\\
        & -2 \sum_f T \int_0^{\infty}  \frac{d^3p}{(2\pi)^3}ln[1+3(\Phi+\overline{\Phi}e^{-(E_{pf}-\mu)/T})e^{-(E_{pf}-\mu)/T})+e^{-3(E_{pf}-\mu)/T})]
        \\
	    & -2 \sum_f T \int_0^{\infty}  \frac{d^3p}{(2\pi)^3}ln[1+3(\overline{\Phi}+{\Phi}e^{-(E_{pf}+\mu)/T})e^{-(E_{pf}+\mu)/T})+e^{-3(E_{pf}+\mu)/T})]
        \\
        &=\Omega-\kappa T^4 ln[J(\Phi,\overline{\Phi})]
    \end{split}
\end{equation}
\end{widetext}
where, $\sigma_f^2 = (\sigma_u^2 + \sigma_d^2 + \sigma_s^2)$ ,$\sigma_f^4 = (\sigma_u^4 + \sigma_d^4 + \sigma_s^4)$ and $\Lambda$ is the three momentum cut off value. As chiral symmetry is broken dynamically, $N_f^{2}-1$ Goldstone bosons do appear. These Goldstone bosons are the pions and kaons. The NJL model parameters have been set using the masses and decay widths of these bosons from experimental findings. The parameter values of the fermionic part of the 2+1 flavor model are: $m_u$=5.5 MeV, $m_s$=134.76 MeV, $\Lambda$=631 MeV, $g_S\Lambda^2$= 3.67 and $g_D\Lambda^5$= 9.33. The pressure of the strongly interacting matter can be written in terms of the potential density as,
\begin{equation}\label{eqn.7}
    P(T,\mu)=-\Omega(T,\mu)
\end{equation}
and the number density of the quarks can be written as,
\begin{equation}\label{eqn.8}
    n_q(T,\mu)=-\frac{\partial{\Omega(T,\mu)}}{\partial{\mu}}
\end{equation}
where T denotes the temperature and $\mu$ is the quark chemical potential.
So far, we have discussed about PNJL model for infinite volume. Now we will discuss its implementation in finite volume. To do this, the first step is to choose the proper boundary condition (Periodic for bosons but anti-periodic for fermions). This would give an infinite sum over the discrete momentum values, given by, $p_i=n_i\pi/R$ (where i=x,y,z and $n_i$ are all positive integers and R is the lateral size of a cubic volume). Then, the effects of surface and curvatures should be incorporated. However, in this calculation, we have taken up some simplifications, such as, (i) neglecting surface and curvature effects, (ii) treating the infinite sum as an integration over a continuous variation of momentum, (iii) not using any modification to the mean field parameters due to finite size effects, as was taken in \cite{deb1}.
%\subsection{\label{sec:level2}Second-level heading: Formatting}
%\subsubsection{Wide text (A level-3 head)}
%\paragraph{Syntax}
%\paragraph{Eliding repeated information}
%\paragraph{The options of the cite command itself}
%{\subsection{Taylor expansion of the pressure}
The freeze-out curve $T(\mu_B)$ in the $T-\mu_B$ plane and the dependence of the baryon chemical potential on the center of mass energy in a nucleus-nucleus collision can be parametrized by \cite{cleymans2006comparison} 
\begin {equation}
T(\mu_B) = a - b\mu_B^2 - c\mu_B^4
\end {equation}
where $a = (0.166 \pm 0.002) $ $GeV$, $b = (0.139 \pm 0.016) $ ${ GeV^{-1}}$,  
$c = (0.053 \pm 0.021) $ $GeV^{-3} $ and
\begin {equation}
\mu_B (\sqrt s_{NN}) = d/{(1+ e\sqrt s_{NN})}
\end {equation}
with $d=1.308\pm 0.028\:GeV$, $e=0.273\pm 0.008\:GeV^{-1}$ \cite{karsch-strange}.
Conclusively, the ratio of baryon to strangeness chemical potential on the freeze-out curve showed a weak dependence on the collision energy.
\begin{equation}
{\mu_S\over\mu_B} \sim 0.164 + 0.018 \sqrt s_{NN}
\end{equation}
Although, several methods for new parametrization for the freeze-out curve in the $T$ vs $\mu_B$ plane have been proposed \textit{e.g.} Borsanyi et al. \cite{borsanyi1},  the old method where the values of baryon, charge, and strangeness chemical potentials are available with respect to freeze out temperature and BES energy has been used \cite{karsch-strange}, so as not to be restricted to the $T-\mu_B$ plane alone. The old method ensures the evaluation of all the values of the chemical potential.\\
\\
The pressure of the strongly interacting matter can be written as,
\begin {equation}
P(T,\mu_B,\mu_Q,\mu_S)=-\Omega (T,\mu_B,\mu_Q,\mu_S),
\label{pres}
\end {equation}

Here,  T  is the temperature,  $\mu_B$  is the baryon (B) chemical potential,  $\mu_Q$  is the charge (Q) chemical potential, and  $\mu_S$  is the strangeness (S) chemical potential. According to thermodynamic relations, the first derivative of pressure with respect to the quark chemical potential  $\mu_q$  yields the quark number density, while the second derivative gives the quark number susceptibility (QNS). By numerically minimizing the thermodynamic potential with respect to the fields  $\sigma_u$,  $\sigma_d$,  $\sigma_s$,  $\Phi$, and  $\bar{\Phi}$, the mean-field value for pressure can be determined (\ref{pres}) %\cite {deb}.

\section{Results}
The influence of density fluctuations on the critical endpoint (CEP) of the QCD phase diagram using the Polyakov-loop enhanced Nambu–Jona-Lasinio (PNJL) model are presented. By examining quark number densities, we aim to identify the critical region and estimate the location of the CEP. Quark number densities, as fundamental thermodynamic quantities, serve as reliable predictors of the CEP, with quark number susceptibilities—derived from their derivatives with respect to the chemical potential—acting as key indicators of critical behavior. We present detailed results on the impact of finite volume effects on quark number density fluctuations, which are recognized as critical signatures of the QCD phase transition. Specifically, we calculate quark number densities and their normalization by $T^3$ as functions of temperature $T$ and $T/\mu_x$, where x represents light quarks, strange quarks, or baryons. By analyzing inflection points and maxima in these curves, we estimate the critical region and pinpoint the CEP’s location within the QCD phase diagram. These findings are compared with experimental data from the RHIC Beam Energy Scan program, providing valuable insights into the behavior of strongly interacting matter near the critical point.

The location of critical point using PNJL model calculation considering both infinite volume and finite volume (with R=2 fm) systems have been discussed. Using PNJL model data sets for the number density of quarks in a system as a function of the temperature of the system at a fixed quark chemical potential ($\mu_q$) for different beam energies (7.7, 11.5, 14.5, 19.6, 27, 39, 62.4, 130, 200 GeV) have been obtained. The quark number density ($n_q$) as a function of temperature (T) and $n_q /T^3$ as a function of temperature (T) is plotted. Using PNJL model parameters similar datasets for strange and baryon numbers densities have been determined. The list of the value of constant $\mu$’s ($\mu_S$ is strange quark number density and $\mu_B$ is the number density of baryon) for different beam energies are shown in Table:\ref{Table:1}
\begin{table}[httb]
\begin{center}
\begin{widetext}
\caption{Value of $\mu_q , \mu_ S$, and $ \mu_B$ for different beam energies}
%\begin{tabular}{  p{1.5cm} p{1.5cm} p{1.5cm} p{1.5cm} p{1.5cm} p{1.5cm} p{1.5cm} p{1.5cm} p{1.5cm} p{1.5cm}}
\begin{tabular}{|c|c|c|c|c|c|c|c|c|c|c|c|}
\hline
 Beam Energy $(\sqrt{s}_{NN})$  & { 7.7 GeV } & { 11.5 GeV } & { 14.5 GeV } & { 19.6 GeV } & { 27 GeV } & { 39 GeV } & { 62.4 GeV } &{ 130 GeV } & { 200 GeV } \\
\hline
 $\mu_q$ (MeV) & 140.33 & 105.33 & 88.00 & 68.67 & 52.00 & 37.33 & 24.33 & 12.00 & 8.00 \\  
 \hline
 $\mu_S$ (MeV) & 127 & 117 & 112 & 106 & 101 & 97 & 94 & 90.14 & 90.34 \\
 \hline
 $\mu_B$ (MeV) & 421 & 316 & 264 & 206 & 156 & 112 & 73 & 36 & 24 \\
 \hline  
\end{tabular}
\label{Table:1}
\end{widetext}
\end{center}
\end{table}

%In the graphs, we have highlighted the temperature points corresponding to the beam energy values used in BES. To get these temperature (T) and $T/\mu_B$ from the value of beam energies ($\sqrt{s_{NN}}$), we have used the following expressions, as was used by Cleymans \cite{cleymans2006comparison}:\\
%\begin{equation}
%T ( \mu_B )=a - b \mu_B^2 - c \mu_B^4
%\label{eq:7}
%\end{equation}
%\begin{equation}
 %\mu_B (\sqrt s)= \frac{d}{1 + e  \sqrt{ s}}
 %\label{eq:8}
%\end{equation}
%\\where, $a = 0.166 \pm 0.002 $GeV, $b = 0.139 \pm 0.016$ GeV$^{-1}$ , $c = 0.053 \pm 0.021$ GeV$^{-3}$ , and
%$d = 1.308 \pm 0.028 $GeV, $e = 0.273 \pm 0.008$ GeV$^{-1}$ .
The critical temperatures corresponding to different collision energies from BES can be obtained by locating the
temperatures at which the light quark chiral condensate has a jump or from the maximal point of the derivative of light quark condensate with respect to the temperature for different chemical potentials.  The maximal point in the derivative of the chiral condensate indicates the critical point from crossover to first order transition for each collision energy. The values of critical temperature, as obtained from calculations, are given in the following Table: \ref{Table:2}
\begin{table}[htb]
\begin{center}  
\begin{widetext}
\caption{Value of $T_c$ for different beam energies}
%\begin{tabular}{  p{1.5cm} p{1.5cm} p{1.5cm} p{1.5cm} p{1.5cm} p{1.5cm} p{1.5cm} p{1.5cm} p{1.5cm} p{1.5cm}}
\begin{tabular}{|c|c|c|c|c|c|c|c|c|c|c|}
\hline
Beam Energy {$(\sqrt{s}_{NN})$} & { 7.7 GeV } & { 11.5 GeV } & { 14.5 GeV } & { 19.6 GeV } & { 27 GeV } & { 39 GeV } & { 62.4 GeV } &{ 130 GeV } & { 200 GeV } \\
\hline
 {Light Quark} (MeV) & 228.70 & 232.15 & 233.50 & 234.65 & 235.50 & 235.95 & 236.20 & 236.25 & 236.25 \\  
 \hline
{Strange Quark} (MeV) & 234.20 & 234.45 & 234.55 & 234.70 & 234.75 & 234.75 & 234.90 & 234.95 & 234.95 \\
 \hline
Baryon (MeV) & 153.15 & 185.50 & 211.75 & 219.40 & 226.50 & 231.25 & 234.15 & 235.65 & 235.90 \\
%& {11.5 GeV}& 232.15 & 234.45 & NA 
 \hline  
\end{tabular}
\label{Table:2}
\end{widetext}
\end{center}
\end{table}
 
\subsection{Strange Quark}
\begin{figure*}[htb]
  \centering
    {{\includegraphics[width=8.6cm]{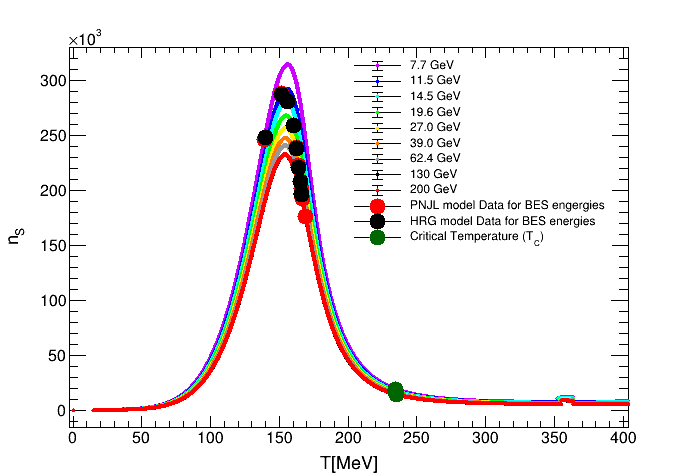} }}
    {{\includegraphics[width=8.6cm]{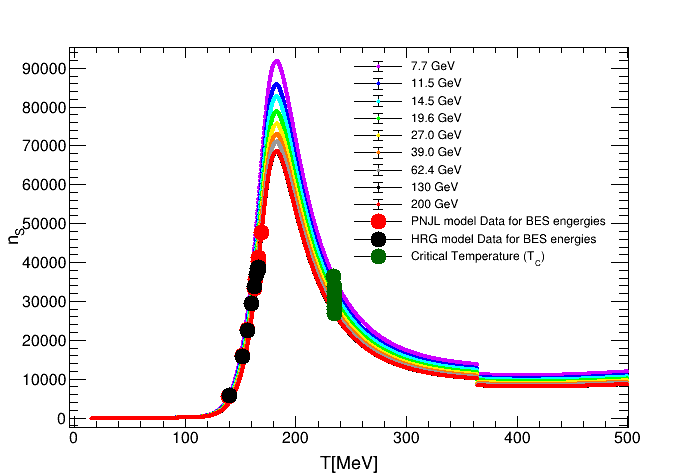} }}
    \caption{(color online) $n_S$ vs. $T$ plot for different beam energies. Left: infinite volume, right: finite volume with R=2 fm. Points correspond to the temperature associated with BES energies from PNJL (red) and HRG (black) model.}
    \label{fig:1}
\end{figure*}

The strange quark number density ($n_S$) as a function of the temperature (T) is shown in Figure.\,\ref{fig:1}. The colored lines in each case, correspond to nine beam energy values (7.7 GeV, 11.5 GeV, 14.5 GeV, 19.6 GeV, 27.0 GeV, 39.0 GeV, 62.4 GeV, 130 GeV, 200 GeV). The $n_S$ vs.T graph has a Poisson-like nature, with its maxima at around T $\sim$ 160 MeV in the case of infinite volume and around T $\sim$ 190 MeV for finite volume case. %Calculation based on PNJL (red point) and on HRG model (black points) are also shown on the corresponding energy curves. The Critical temperature ($T_c$) calculation based on PNJL model are shown in green points and compared. 
%\begin{center}
%\begin{widetext}
%\end{widetext}
%\end{center}
\\
Now, at the two sides of the maxima, the first order derivative or the slope of these curves (positive or negative) has maxima. The region between these two points is the critical region. So, we can expect that the critical end point lies near the maxima of these graphs. The green points refer to the values of critical temperature for each beam energy obtained from theoretical calculations. For each line, the green points are at the base of the Poisson curve. The red points correspond to the beam energy points obtained using Eq.\,(\ref{eqn.7}) and Eq.\,(\ref{eqn.8}). The black points refer to similar calculations, corresponding to the HRG model \cite{hrg_data}. For the infinite volume case, both the red and black points lie near the maxima of the lines. But for the finite volume case, the points lie far left from the maxima. PNJL and HRG model based calculations for BES energies are close to each other and the calculated $T_c$ values are close to 235 MeV for both the finite and infinite cases.

\begin{figure}[htb]
    \centering
%  \subfloat{{\includegraphics[width=8.6cm]{NByT3VsT_6S.png} }}
    \subfigure{{\includegraphics[width=8.6cm]{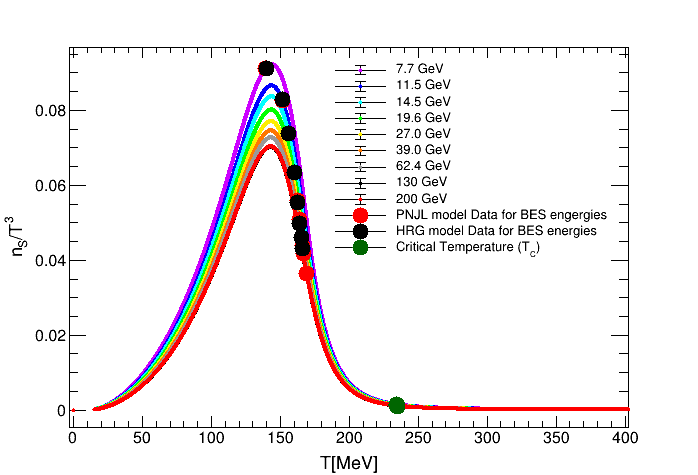} }}
%    \qquad
    \subfigure{{\includegraphics[width=8.6cm]{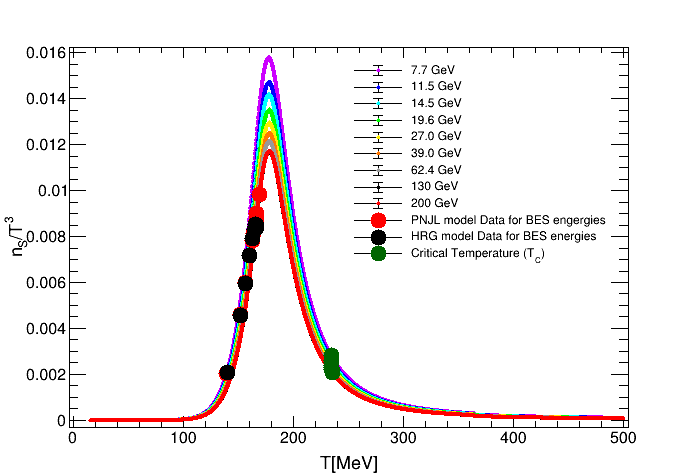} }}
 %   \subfloat{{\includegraphics[width=8.6cm]{NByT3VsT_2S.png} }}
    \caption{(color online) $n_S/T^3$ vs. $T$ for different beam energies. Left: infinite volume, right: finite volume with R=2 fm. Points correspond to the temperature associated with BES energies from PNJL (red) and HRG (black) model.}
    \label{fig:2}
\end{figure}

The $n_S/T^3$ vs. T graphs in Figure.\,\ref{fig:2} have the same Poisson-like nature. So, the nature of the slope of these curves, changes at the neighborhood of the maxima. Therefore, we can expect that the critical point is present near the maxima of these curves too. In the case of infinite volume, for lower beam energy (7.7 GeV, 11.5 GeV), the red and black points lie near the maxima of the lines. But, as the energy value increases, the points shift to the right of the maxima more and more. For finite volume case, these points are at the left of the maxima for each line (energy). The critical temperature points (green) lie near the base of the lines, for each case.
%\begin{center}
%\begin{widetext}
\begin{figure*}[htb]
    \centering
    {{\includegraphics[width=8cm]{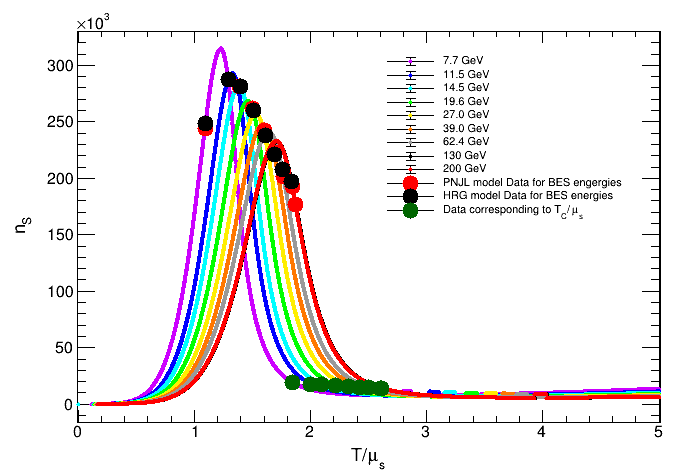} }}%
    %\qquad
    {{\includegraphics[width=8cm]{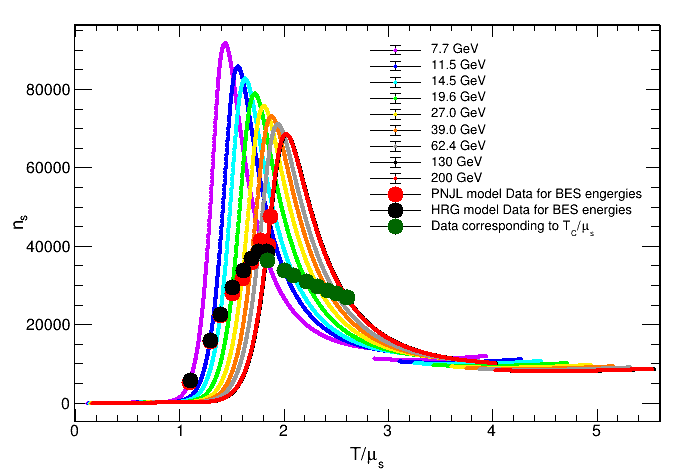} }}%
    \caption{(color online) $n_S$ vs. $T/\mu_s$ plot for different beam energies. Left: infinite volume, right: finite volume with R=2 fm. Points correspond to the temperature associated with BES energies from PNJL (red) and HRG (black) model.}%
    \label{fig:3}%
\end{figure*}
%\end{widetext}
%\end{center}

The x-axis (temperature axis) of the $n_S$ vs. T graphs can be scaled by diving it by the value of constant $\mu_S$ for each beam energies to get $n_S$ vs. $T/ \mu_S$ plots in Figure.\,\ref{fig:3}. From the $n_S$ vs. $T/ \mu_S$ plots, the peaks for different beam energies can be seen clearly. From these peaks, we can estimate the temperature at which, the critical end point may present. Red and black points lie near the maxima for lower beam energies in case of infinite volume, but far away from the peaks for finite volume system in all energies.
%\begin{center}
%\begin{widetext}
\begin{figure*}[htb]
    \centering
    {{\includegraphics[width=8cm]{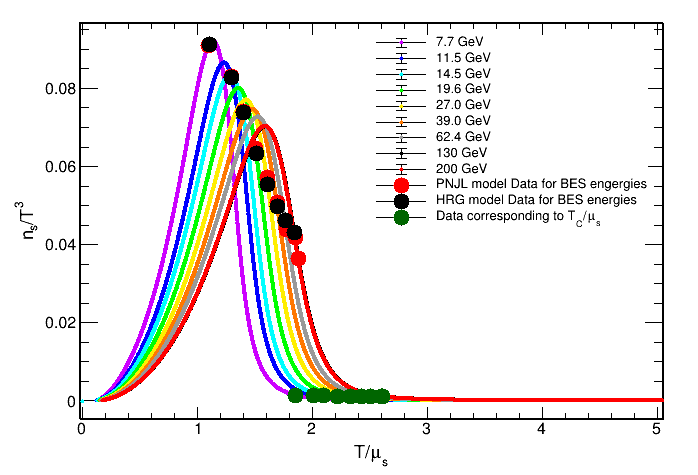} }}%
    %\qquad
    {{\includegraphics[width=8cm]{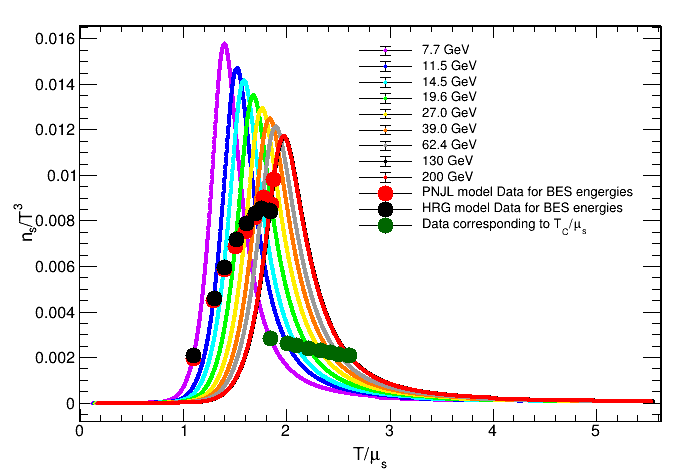} }}%
    \caption{(color online) $n_S/T^3$ vs. $T/\mu_s$ plot for different beam energies. Left: infinite volume, right: finite volume with R=2 fm. Points correspond to the temperature associated with BES energies from PNJL (red) and HRG (black) model.}%
    \label{fig:4}%
\end{figure*}
%\end{widetext}
%\end{center}

The $n_S /T^3$ vs. $T/\mu_S$ plots in Figure.\,\ref{fig:4} have the same nature that of $n_S$ vs. $T/ \mu_S$ plots. So, similar type of conclusions can be made from these graphs too. The position of the three types of points are similar to that of above plots.

\subsection{Light Quark}
\begin{figure*}[htb]%
    \centering
    {{\includegraphics[width=8cm]{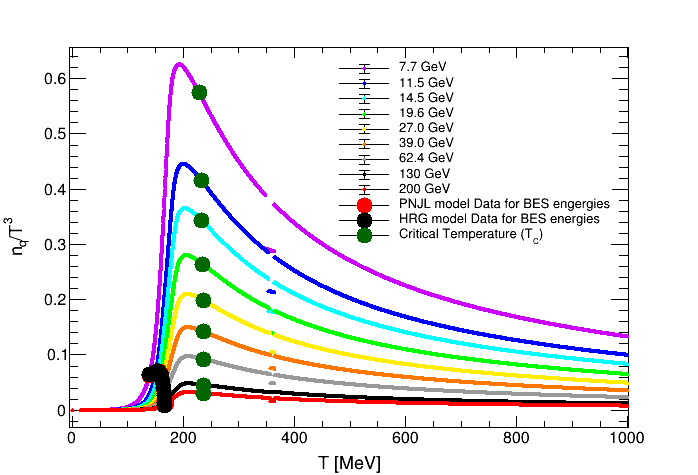} }}%
    %\qquad
    {{\includegraphics[width=8cm]{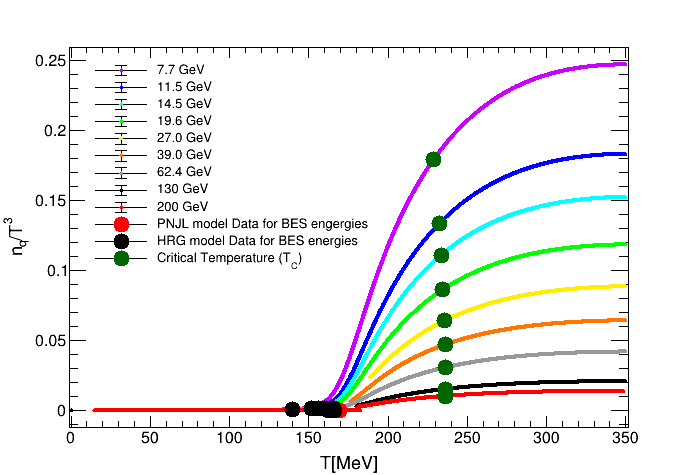} }}%
    \caption{(color online) $n_q/T^3$ vs. $T$ plot for different beam energies. Left: infinite volume, right: finite volume with R=2 fm. Points correspond to the temperature associated with BES energies from PNJL (red) and HRG (black) model.}%
    \label{fig:5}%
\end{figure*}

The variations of $n_q /T^3$ as a function of $T$ for light quark for finite and infinite volume are shown in Figure.\,\ref{fig:5}. This $n_q /T^3$ vs. T plots [\ref{fig:5}] do not have a Poisson-like nature. These plots have a rapid change in their slope, in the small T region and the slope decreases slowly after the peak for the infinite volume case. An inflection point is located near the left of the peak, and another one is a little bit further from the peak at the right. So the critical point is assumed to be at the peak of the curves. The red and black points are very far away (at left) from the peak for every energy value. The $T_{C}$ points (green points) are located very near the peak of the curves. In the case of finite volume system size, the peaks are not present (may be due to unavailability of data). Both the calculated $T_c$ values are similar for finite and infinite cases.
%\begin{center}
%\begin{widetext}
%\end{widetext}
%\end{center}
\par
\begin{figure*}[htb]%
    \centering
    {{\includegraphics[width=8cm]{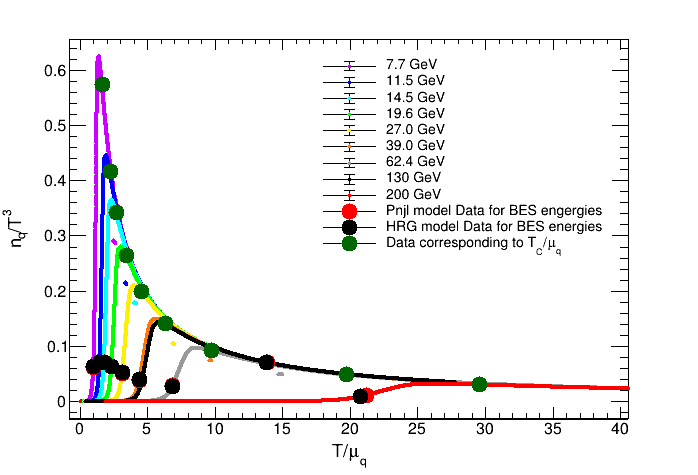} }}%
    %\qquad
    {{\includegraphics[width=8cm]{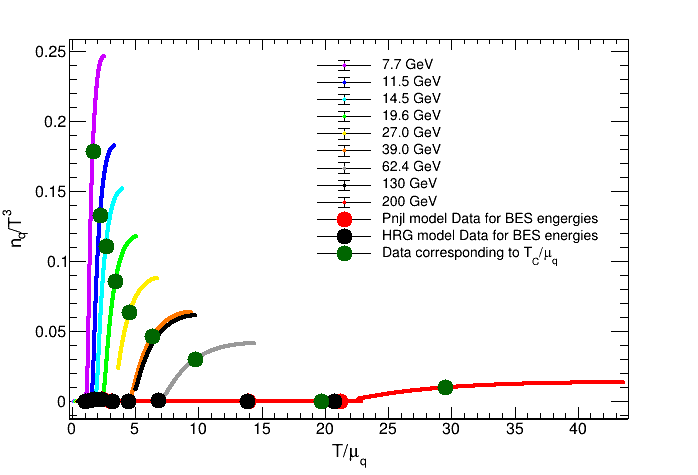} }}%
    \caption{(color online) $n_q/T^3$ vs. $T/\mu_q$ plot for different beam energies. Left: infinite volume, right: finite volume with R=2 fm. Points correspond to the temperature associated with BES energies from PNJL (red) and HRG (black) model.}%
    \label{fig:6}%
\end{figure*}

The $n_q /T^3$ vs. $T/\mu_q$ plots in Figure.\,\ref{fig:6} have the same kind of nature as the $n_q /T^3$ vs. T graphs, as the only difference is that, the T axis is scaled by a factor of $\mu_q$ . So, the position of the CEP can also be estimated from these graphs. The calculated $T_c$ values (green points) are very near to the peak of the curves. The other two types of points (points corresponding to the temperature associated with BES energies, calculated in PNJL and HRG model) are far away from the peak.
%\begin{center}
%\begin{widetext}
%\end{widetext}
%\end{center}

\begin{figure*}[htb]%
    \centering
    {{\includegraphics[width=8cm]{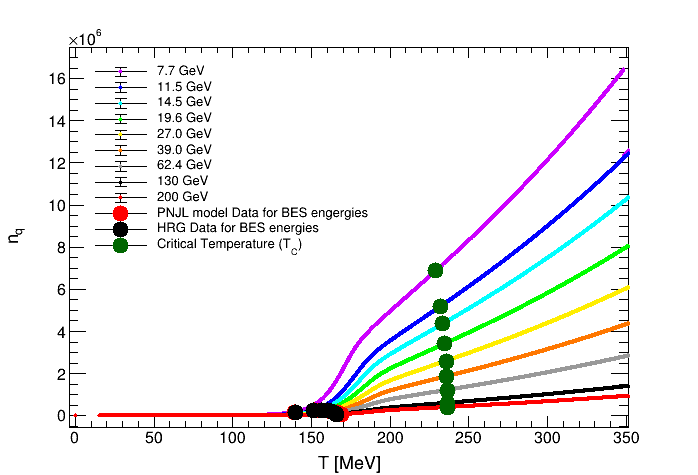} }}%
    %\qquad
    {{\includegraphics[width=8cm]{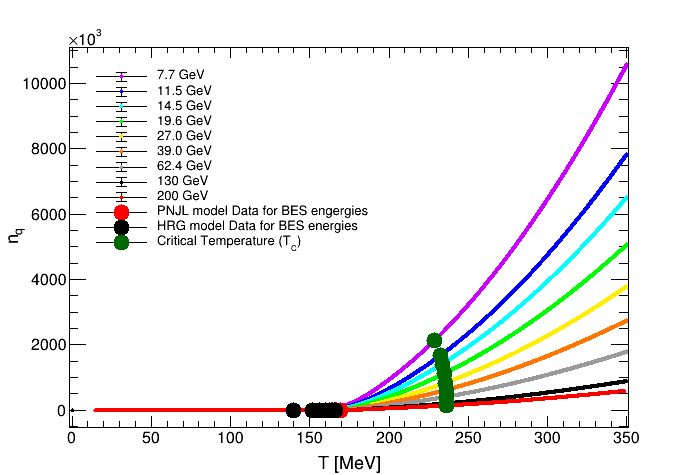} }}%
    \caption{(color online) $n_q$ vs. $T$ plot for different beam energies. Left: infinite volume, right: finite volume with R=2 fm. Points correspond to the temperature associated with BES energies from PNJL (red) and HRG (black) model.}%
    \label{fig:7}%
\end{figure*}

\begin{figure*}[htb]%
    \centering
    {{\includegraphics[width=8cm]{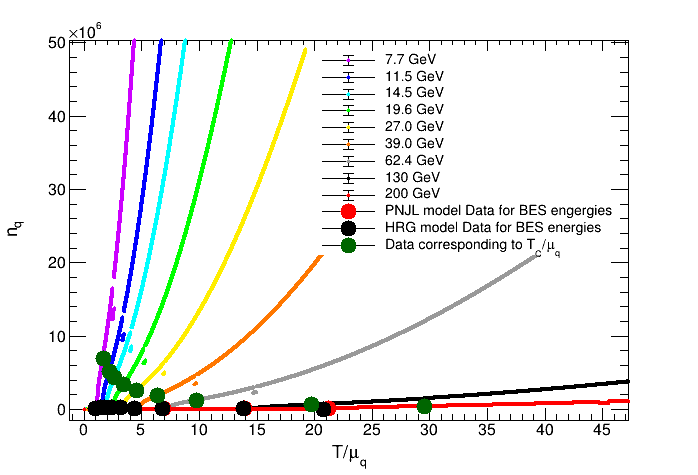} }}%
    %\qquad
    {{\includegraphics[width=8cm]{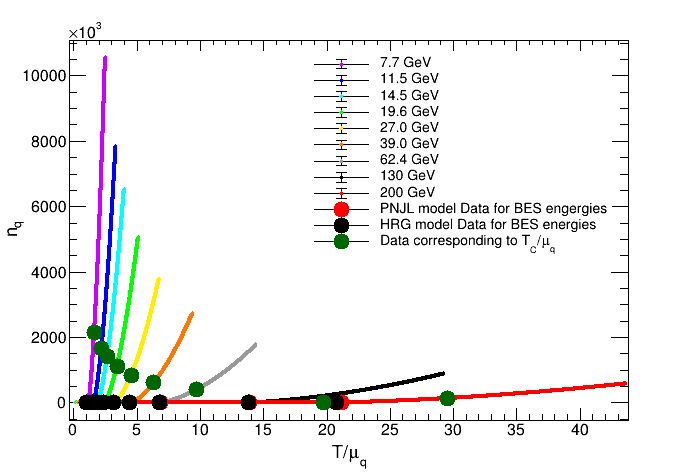} }}%
    \caption{(color online) $n_q$ vs. $T/\mu_q$ plot for different beam energies. Left: infinite volume, right: finite volume with R=2 fm. Points correspond to the temperature associated with BES energies from PNJL (red) and HRG (black) model.}%
    \label{fig:8}%
\end{figure*}

The variations of $n_q $ as a function of $T$ for light quark for finite and infinite volume are shown in Figure.\,\ref{fig:7}. The $n_q$ vs. T graphs have no maxima as such for both the finite and infinite volume. Finding CEP from this type of plot is difficult. In the region between the black and green points, there is an inflection point, where the slope of the curves have changed it's nature. Critical point lies near that point.
%\begin{center}
%\begin{widetext}
%\end{widetext}
%\end{center}
Figure.\,\ref{fig:8} shows the variation of $n_q$ as a function of the temperature scaled with $\mu_q$ for finite and infinite volume. The $n_q$ vs. $T/\mu_q$ curves for different beam energies have a clear point of inflection for the infinite volume case. For finite volume case, this point is not so clear. The green points lie at the right, and the black (and red) points lie at the left of that point. The HRG and PNJL calculations are in good agreement for all beam energies.
%\begin{center}
%\begin{widetext}
%\end{widetext}
%\end{center}
\subsection{Baryon}
\begin{figure*}[htb]%
    \centering
    {{\includegraphics[width=8cm]{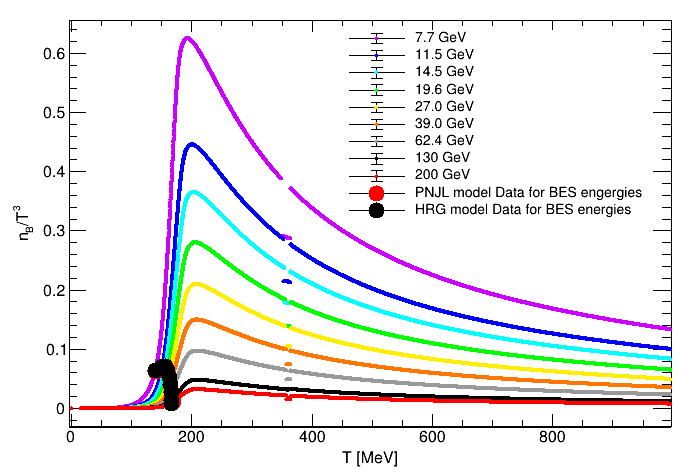} }}%
    %\qquad
    {{\includegraphics[width=8cm]{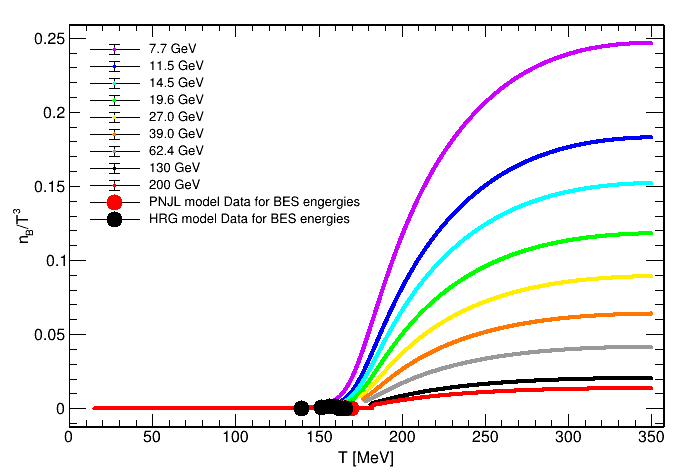} }}%
    \caption{(color online) $n_B/T^3$ vs. $T$ plot for different beam energies. Left: infinite volume, right: finite volume with R=2 fm. Points correspond to the temperature associated with BES energies from PNJL (red) and HRG (black) model.}%
    \label{fig:9}%
\end{figure*}

For baryon number density the variations of $n_B /T^3$ as a function of $T$ for infinite and finite volume are shown in Figure.\,\ref{fig:9} for different beam energies. The variations has similar dependency as in quark number density. These curves have a rapid change in their slope, in the small T region and the slope decreases slowly after the peak for the infinite volume case. An inflection point is located near the left of the peak, and another one is a little bit further from the peak at the right. So the critical point is assumed to be at the peak of the curves. The red and black points are very far away (at left) from the peak for every energy value. Calculations based on PNJL (red point) and on HRG model (black points) are also shown on the corresponding energy curves. 
%\begin{center}
%\begin{widetext}

\begin{figure*}[htb]%
    \centering
    {{\includegraphics[width=8cm]{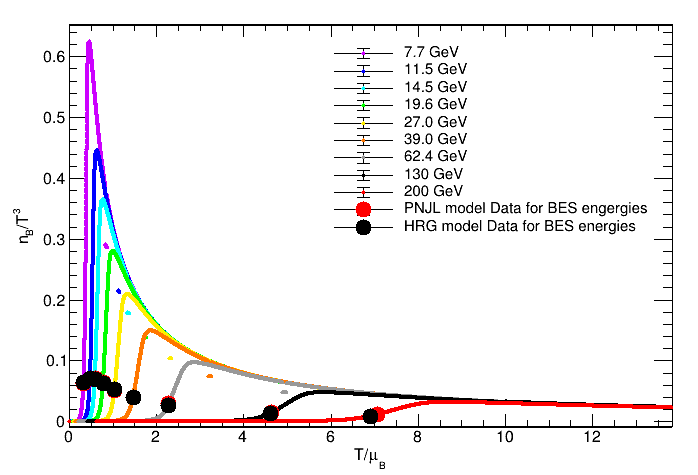} }}%
    %\qquad
    {{\includegraphics[width=8cm]{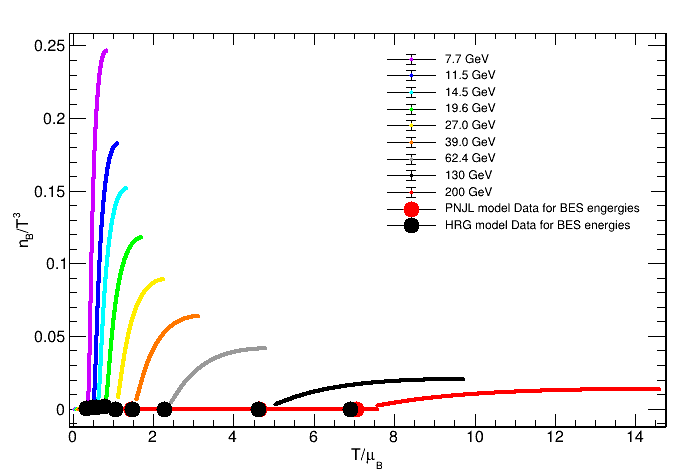} }}%
    \caption{(color online) $n_B/T^3$ vs. $T/\mu_B$ plot for different beam energies. Left: infinite volume, right: finite volume with R=2 fm. Points correspond to the temperature associated with BES energies from
PNJL and HRG model.}%
    \label{fig:10}%
\end{figure*}

Figure.\,\ref{fig:10} shows $n_B /T^3$ as a function of $T/\mu_B$ for different RHIC energies for infinite and finite volume. The position of the CEP can also be estimated from these graphs. The points corresponding to temperatures associated with BES energies, calculated in PNJL and HRG model are also shown and are far away from the peak.
%The calculated $T_c$ values (green points) are very near to the peak of the curves.

\begin{figure*}[htb]%
    \centering
    {{\includegraphics[width=8cm]{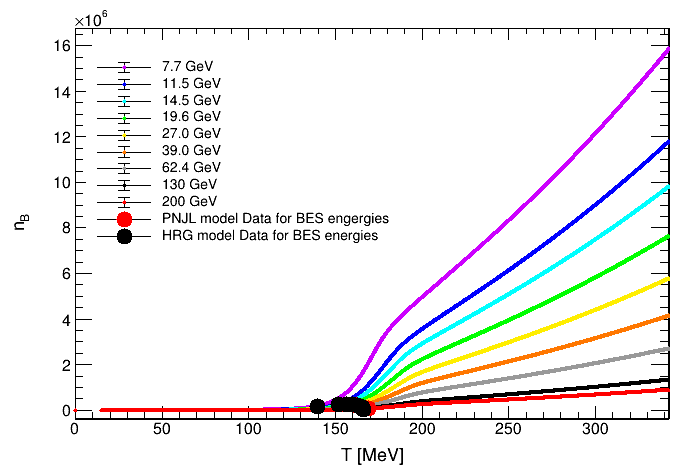} }}%
    %\qquad
    {{\includegraphics[width=8cm]{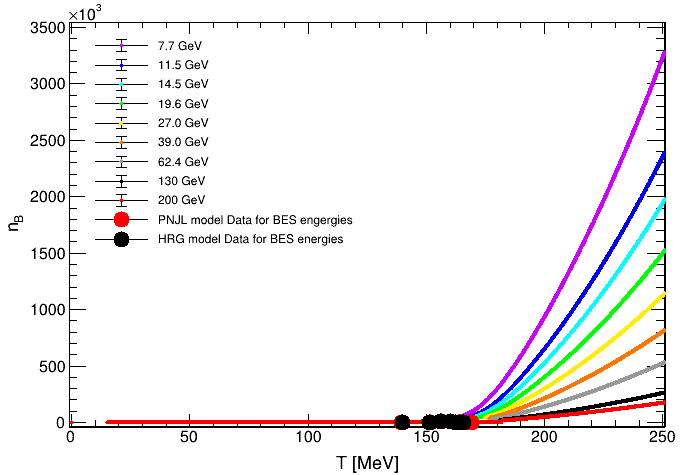} }}%
    \caption{(color online) $n_B$ vs. $T$ plot for different beam energies. Left: infinite volume, right: finite volume with R=2 fm. Points correspond to the temperature associated with BES energies from PNJL (red) and HRG (black) model.}%
    \label{fig:11}%
\end{figure*}

%\end{widetext}
%\end{center}
The variations of $n_B$ as a function of $T$ for different energies for infinite and finite volume are shown in Figure.\,\ref{fig:11}. Here also, the variations has similar dependency as in quark number density. These curves have no maxima as such for both the finite and infinite volume. Finding CEP from this type of plot is difficult.

%\begin{center}
%\begin{widetext}
\begin{figure*}[htb]%
    \centering
    {{\includegraphics[width=8cm]{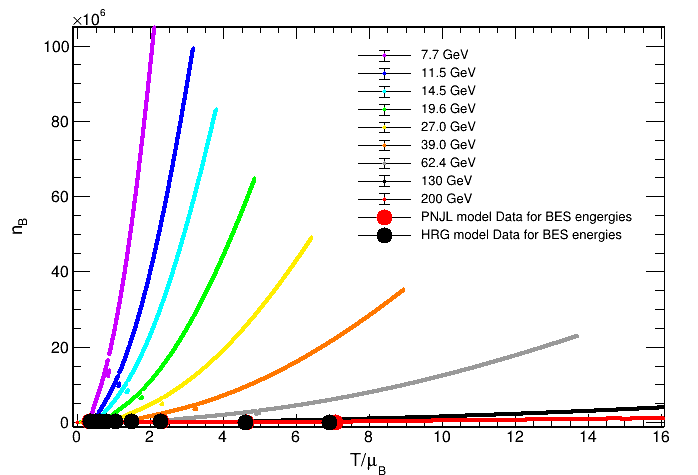} }}%
    %\qquad
    {{\includegraphics[width=8cm]{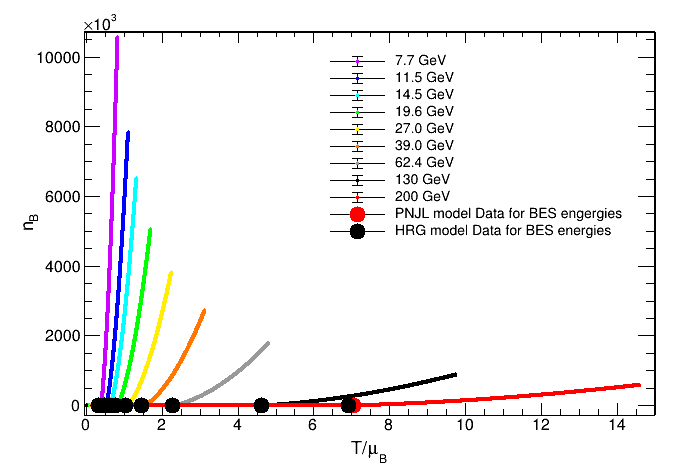} }}%
    \caption{(color online) $n_B$ vs. $T/\mu_B$ plot for different beam energies. Left: infinite volume, right: finite volume with R=2 fm. Points correspond to the temperature associated with BES energies from PNJL (red) and HRG (black) model.}%
    \label{fig:12}%
\end{figure*}

%\end{widetext}
%\end{center}
The variations of $n_B$ as a function of $T/\mu_B$ for different energies for infinite and finite volume are shown in Figure.\,\ref{fig:12}. The $n_B$ vs. $T/\mu_B$ curves for different beam energies have a clear point of inflection for the infinite volume case. For finite volume case, this point is not so clear. The  black (and red) points lie at the left of that point. The HRG and PNJL calculations are in good agreement for all beam energies. All of these four kinds of plots for the baryons show the exactly same nature as the quarks. So, the location of the critical points in the QCD phase diagram can be estimated from these plots similarly.
%\begin{center}
%\begin{widetext}
%\end{widetext}
%\end{center}
\\\\

\section{Summary}

%In this work, effort has been made to find out the critical end point of the QCD phase diagram using the PNJL model.% This is a very important work at hand, for high energy physicists, all over the world. 
%We have discussed the three flavors PNJL model for infinite and finite volume systems. The calculation of the quark number densities for light quarks and strange quarks for different energies of the infinite and finite volume system has been performed. The number density of different systems has been presented as a function of energy comparable to the RHIC Beam Energy Scan. The nature of the quark number densities as a function of energy have been analyzed. From the behavior of the data for the finite and infinite PNJL systems, the existence of the critical point has been discussed.
This work offers deeper insights into the phase structure of QCD and contributes to the ongoing search for the elusive critical endpoint in high-energy heavy-ion collisions. By highlighting the role of finite volume effects, it bridges theoretical predictions and experimental observations, advancing our understanding of the QCD phase transition.
In this work, an effort has been made to identify the critical end point of the QCD phase diagram using the PNJL model. This study holds significant importance for high-energy physicists worldwide. We have discussed the three-flavor PNJL model for both infinite and finite volume systems. The quark number densities for light and strange quarks were calculated for various energies in both infinite and finite volume systems. These number densities have been presented as a function of energy, comparable to the RHIC Beam Energy Scan. The behavior of the quark number densities as a function of energy has been analyzed. Based on the data for the finite and infinite PNJL systems, the existence of the critical point has been examined.

To conclude, we have studied the finite volume number density fluctuations at energies similar to RHIC BES with system sizes both finite and infinite in the PNJL model. The variation of $n_q/T^3$ with $T$ has the same general behavior in both cases. The peak value for $n_q /T^3$ is around 0.44 for infinite volume and around 0.18 for finite volume at $\mu = 105.33$ MeV (corresponding to $\sqrt{s}_{NN} = 11.5$ GeV), which is lower than the former. A similar difference in peak values of $n_q/T^3$ is observed for other beam energies as well. This trend is also seen in other number distributions. Therefore, we can conclude that the quark and baryon number densities depend on the system size.

The nature of our $n_S$ vs. $T$ (or $T/\mu_q$) and $n_S/T^3$ vs. $T$ (or $T/\mu_q$) graphs for strange quarks resembles Poisson distributions, showing points of inflection on both sides of the maxima. We have highlighted the temperature points corresponding to BES energies on the curves, some of which lie at the maxima, while others fall on either side between the inflection point and the maxima. These points do not at the $T_c$ values, indicating that the exact location of the critical endpoint may not precisely match the measured RHIC beam energy values. However, this study suggests that RHIC measurements are very close to the point of interest. Therefore, at the energies where Beam Energy Scans are conducted, we may not observe all the behaviors that occur at the critical point. Although these measured points are located near the desired region, the measurements will be affected by the influence of the critical region, and the area near the critical endpoint can be explored by RHIC BES. Further studies of the BES at intermediate energies with higher statistics are required to determine the exact location of the QCD critical point.

\begin{acknowledgments}
P.Deb would like to thank Women Scientist Scheme A (WOS-A) of the Department of Science and Technology (DST) funding with grant no SR/WOS-A/PM-10/2019 (GEN). 
\end{acknowledgments}


%Beam \\Energy\\ $(\sqrt{s}_{NN})$} & {7.7\\GeV} & {11.5\\GeV} & {14.5\\GeV} & {19.6\\GeV} & {27.0\\GeV} & {39.0\\GeV} & {62.4\\GeV} &{130.0\\GeV} & {200.0\\GeV} \\
%\makecell[l]{ Beam \\Energy\\ $(\sqrt{s}_{NN})$} & \makecell[l]{7.7\\GeV} & \makecell[l]{11.5\\GeV} & \makecell[l]{14.5\\GeV} & \makecell[l]{19.6\\GeV} & \makecell[l]{27.0\\GeV} & \makecell[l]{39.0\\GeV} & \makecell[l]{62.4\\GeV} & \makecell[l]{130.0\\GeV} & \makecell[l]{200.0\\GeV} \\

%\nocite{*}
%\bibliographystyle{plain}
%\bibliographystyle{plain} 
%\bibliography{book}
%\bibliography{book}% Produces the bibliography via BibTeX.
%\bibliographystyle{ieeetr}
%\bibliography{book}
%\bibliographystyle{book}
%\printbibliography

\end{document}